\documentclass[a4paper,11pt]{article}
\usepackage{pos}
\usepackage{amsmath}    

\title{Studies of nucleon isovector structure with the PACS10 superfine lattice}
\ShortTitle{Studies of nucleon isovector structure with the PACS10 superfine lattice}

\author*[\dag, a]{Ryutaro Tsuji}
\author[b]{Yasumichi Aoki}
\author[c]{Ken-Ichi Ishikawa}
\author[d]{Yoshinobu Kuramashi}
\author[e]{Shoichi Sasaki}
\author[f]{Kohei Sato}
\author[d]{Eigo Shintani}
\author[g]{Hiromasa Watanabe}
\author[h.d]{Takeshi Yamazaki}
\affiliation[]{\normalsize{\bf \sffamily \hspace{50mm}(PACS Collaboration)}}

\affiliation[a]{High Energy Accelerator Research Organization (KEK),
  305-0801, Tsukuba, Japan}

\affiliation[b]{RIKEN Center for Computational Science,
  650-0047, Kobe, Japan}

\affiliation[c]{Core of Research for the Energetic Universe,
    Graduate School of Advanced Science and Engineering, Hiroshima University,
  739-8526, Higashi-Hiroshima, Japan}

\affiliation[d]{Center for Computational Sciences, University of Tsukuba,
  305-8577, Tsukuba, Japan}

\affiliation[e]{Department of Physics, Tohoku University,
  980-8578,Sendai, Japan}

\affiliation[f]{Degree Programs in Pure and Applied Sciences, Graduate School of Science and Technology, //
University of Tsukuba, Ibaraki 305-8571, Japan}

\affiliation[g]{Yukawa Institute for Theoretical Physics, Kyoto University, Kyoto 606-8502, Japan}

\affiliation[h]{Institute of pure and Applied Sciences, University of Tsukuba,\\
  305-8571, Tsukuba, Japan}

\emailAdd{rtsuji@post.kek.jp}

\abstract{
    We present the results for the nucleon axial-vector, induced pseudoscalar and pion-nucleon couplings obtained from 2+1 flavor lattice QCD at the physical point with a large spatial extent of about 10 fm.
    Our calculations are performed with the PACS10 gauge configurations generated by the PACS Collaboration with the six stout-smeared $O(a)$ improved Wilson-clover quark action and Iwasaki gauge action at $\beta$ = 1.82, 2.00 and 2.20 corresponding to lattice spacings of 0.09 fm (coarse), 0.06 fm (fine) and 0.04 fm (superfine), respectively. 
    We first evaluate the value of the nucleon axial-vector coupling.
    In addition,
    the induced pseudoscalar and pion-nucleon couplings from the induced pseudoscalar form factor are also investigated.
    Combining the results obtained from the all of our coarse, fine and superfine lattices,
    we finally discuss the systematic uncertainties in our calculation
    based on the comparison with both of the experimental values and lattice QCD results provided by the other collaborations.
}

\FullConference{The 41st International Symposium on Lattice Field Theory (LATTICE2024)\\
 28 July - 3 August 2024\\
Liverpool, UK\\}

\begin{document}
\maketitle

\section{Introduction}
\label{sec:introduction}
In the standard model of modern particle physics, protons and neutrons, known as nucleons, are composite particles of quarks and gluons, and the interaction among them is formulated as Quantum Chromodynamics (QCD).
This indicates that the nucleon has a non-trivial structure due to the complex dynamics of QCD.
Especially, the axial structure of the nucleon plays a relevant role in the description of the experiments involving the weak interaction.
Indeed,
the axial and induced pseudoscalar form factor which describe the nucleon-neutrino scattering are especially desired to be studied,
because they provide important information for current neutrino oscillation experiments such as
T2K by Super- and Hyper-K~\cite{T2K:2011qtm, Hyper-KamiokandeProto-:2015xww, T2K:2019bcf}, or NOvA by DUNE~\cite{NOvA:2019cyt, DUNE:2020ypp}.

The nucleon axial-vector coupling $g_A$, 
which
is associated with the neutron lifetime puzzle~\cite{Czarnecki:2018okw},
can be determined from the axial form factor at zero-momentum transfer or
by the $\beta$-decay measurements with cold and ultracold neutrons.
Indeed,
this coupling is the most precisely known quantity evaluated as $g_A=1.2756(13)$~\cite{ParticleDataGroup:2022pth} for the axial structure of nucleon.
In contrast,
the induced pseudoscalar coupling $g_P^*$ and pion-nucleon coupling $g_{\pi NN}$ defined through the induced pseudoscalar form factor at some finite momentum transfer $q^2$ are much less well known than $g_A$.
The couplings of the induced pseudoscalar and pion-nucleoncoupling obtained by the experiments of 
the muon capture experiments~\cite{MuCap:2015boo}
and the pion-nucleon scattering~\cite{Babenko:2016idp, Limkaisang:2001yz}
with some help of the theoretical analysis using 
the low-energy chiral lagrangian or models.
From the theoretical side,
much effort has been devoted to the high-precision determination of the three couplings using
lattice QCD calculations~\cite{FlavourLatticeAveragingGroupFLAG:2021npn}.

In this work, we use lattice QCD which is formulated in discretized Euclidean space-time,
in order to rigorously solve the low-energy physics of QCD numerically with High Performance Computing (HPC).
Lattice QCD is the only {\it ab initio} method that can numerically calculate quark-gluon dynamics, and thus lattice QCD is essential for studying the internal structure of the nucleon in terms of quark and gluon degrees of freedom.
In general, the observables determined from lattice QCD simulations are subject to both statistical and systematic uncertainties. 
As for the systematic uncertainties on the couplings that can be determined from the corresponding form factors, it is known that there are four major sources: 1) chiral extrapolation, 2) finite size effect, 3) finite lattice spacing effect and 4) excited-state contamination.
As for the nucleon axial-vector coupling $g_A$,
indeed, 
lattice QCD successfully reproduces the experimental values of the axial-vector coupling with high-accuracy 
(see the recent studies in Ref.~\cite{FlavourLatticeAveragingGroupFLAG:2021npn, Bali:2023sdi, Djukanovic:2022wru, Tsuji:2022ric, Park:2021ypf})
by handling those systematic uncertainties,
though it is only an important benchmark study for the structure of the nucleon based on lattice QCD. 
On the other hand,
although great efforts have been made to improve our knowledge of the nucleon,
it is not accurate enough to solve problems (e.g. the proton radius puzzle~\cite{Pohl:2010zza}) or to provide a high-precision input for current neutrino oscillation experiments~\cite{Tomalak:2023pdi}.
Especially for the axial structure,
the leading excited-contamination of $\pi N$ state has been a long-standing obstacle to precisely determine the form factors from lattice QCD~\cite{Gupta:2024krt}.

In this work,
we present the result of our preliminary results for the three nucleon isovector couplings related to the nucleon axial structure with PACS10 superfine lattice.
In addition,
our previous works with the PACS10 coarse~\cite{Shintani:2018ozy} and fine~\cite{Tsuji:2023llh} lattices are also presented to investigate the lattice spacing effect,
which has not yet been examined in our study.

\section{Method}
\label{sec:method}
In this work,
we calculate the axial-vector coupling $g_A$, the induced pseudoscalar coupling $g_P^*$ and the pion-nucleon coupling $g_{\pi NN}$ by using lattice QCD.
These quantities are defined through the corresponding form factors of the axial ($F_A$) and the induced pseudoscalar ($F_P$) as
\begin{align}
    \label{eq:couplings}
   g_A=F_A(q^2=0),\quad g_P^* = m_\mu F_P(q^2=0.88m_\mu^2),\ \mathrm{and}\quad
   g_{\pi NN} \equiv \lim_{q^2 \to -m_\pi^2}\frac{m_\pi^2 + q^2}{2F_\pi}F_{P}(q^2)
\end{align}
with muon mass $m_\mu$ and pion mass $m_\pi$.

The two form factors appear in the description of the standard $\beta$-decay.
Let us consider, for example,
the nucleon matrix elements of the axial-vector current which is included in the weak current that describes the process.
The nucleon matrix element of a renormalized quark bilinear operator~\footnote{
In this paper, we use the notation of Ref.~\cite{Tsuji:2023llh} for the description of the bare and renormalized quantities.
}
$A_\alpha= \bar{u}\gamma_\alpha \gamma_5 d$
is decomposed into the following relativistically covariant forms in terms of the
$F_A$ and $F_P$:
\begin{align}
    \label{eq:nme_fap}
    \langle p(\boldsymbol{p})| A_\alpha(x) | n(\boldsymbol{p}^{\prime}) \rangle 
    =
    \bar{u}_p(\boldsymbol{p})
    \left(
    \gamma_\alpha\gamma_5 F_A(q^2) + iq_\alpha\gamma_5 F_P(q^2)
    \right)
    u_n(\boldsymbol{p}^{\prime}) e^{iq\cdot x},
\end{align}
where $|p(\boldsymbol{p})\rangle$ and $|n(\boldsymbol{p})\rangle$ are the proton ($p$) and neutron ($n$) ground states with the three-dimensional momentum $\boldsymbol{p}$.
In the above equation, the four-dimensional momentum transfer $q$ between the proton and neutron is given by $q=P^\prime-P$ with
$P=(E_p(\boldsymbol{p}), \boldsymbol{p})$ and $P^\prime=(E_n(\boldsymbol{p}^\prime), \boldsymbol{p}^\prime)$.
It is important to note that this matrix element is 
flavor changing process,
where there is no disconnected contribution.

For the lattice QCD computation~\footnote{
In this study, 
the exponentially smeared quark operator $q_S(t,\boldsymbol{x})=\sum_{\boldsymbol{y}}A\mathrm{e}^{-B|\boldsymbol{x}-\boldsymbol{y}|}q(t,\boldsymbol{y})$
with the Coulomb gauge fixing is used for the construction of the nucleon interpolating operator as well as a local quark operator $q(t,\boldsymbol{x})$.
For detail, see Ref.~\cite{Tsuji:2023llh} and references therein. 
}
,
the nucleon two-point function with nucleon interpolating operator located at smeared (\textit{S}) or local (\textit{L}) source $(t_{\rm src})$, and local sink $(t_{\rm sink})$ is constructed as
\begin{align}
    \label{eq:two_pt_func}
    C_{XS}(t_{\mathrm{sink}}-t_{\mathrm{src}}; \boldsymbol{p})
    =
    \frac{1}{4}\mathrm{Tr}
    \left\{
        \mathcal{P_{+}}\langle N_X(t_{\mathrm{sink}}; \boldsymbol{p})
        \overline{N}_S(t_{\mathrm{src}}; -\boldsymbol{p})  \rangle
    \right\}\ \mathrm{with}\ X=\{S,L\},
\end{align}
where $\mathcal{P}_+=(1+\gamma_4)/2$ is the projection operator for the positive parity nucleon.
The nucleon operator with a three-dimensional momentum $\boldsymbol{p}$ is given for the proton state by
\begin{align}
    \label{eq:interpolating_op}
N_L(t,\boldsymbol{p})&
    =\sum_{\boldsymbol{x}}e^{-i\boldsymbol{p}\cdot\boldsymbol{x}}\varepsilon_{abc}
    \left[
        u^{T}_{a}(t,\boldsymbol{x})C\gamma_5d_b(t,\boldsymbol{x})
    \right]
    u_c(t,\boldsymbol{x})
\end{align}
with the charge conjugation matrix, $C=\gamma_4\gamma_2$. The superscript $T$ denotes a transposition, while the indices $a$, $b$, $c$ and
$u$, $d$ label the color and the flavor, respectively.
In addition, 
we evaluate the nucleon three-point functions, which are constructed with the spatially smeared source and
sink operators of the nucleon as 
\begin{align}
    \label{eq:three_pt_func}
    C^{5z}_{A_\alpha}
    (t; \boldsymbol{p}^{\prime}, \boldsymbol{p})
    =
    \frac{1}{4} \mathrm{Tr}
    \left\{
        \mathcal{P}_{5z}\langle N_S(t_{\mathrm{sink}}; \boldsymbol{p}^{\prime})
        A_{\alpha}
        (t; \boldsymbol{q}=\boldsymbol{p}-\boldsymbol{p}^{\prime})
        \overline{N}_S(t_{\mathrm{src}}; -\boldsymbol{p})  \rangle
    \right\}
\end{align}
with
$\mathcal{P}_{5z} = \mathcal{P}_{+}\gamma_5\gamma_z$ which is the projection operator to
extract the form factors.

To extract the form factors,
we take an appropriate combination of two-point function of Eq.~(\ref{eq:two_pt_func}) and three-point function of Eq.~(\ref{eq:three_pt_func}), 
\begin{align}
\mathcal{R}^{5z}_{A_{\alpha}}
\left(t; \boldsymbol{p}^{\prime}, \boldsymbol{p}\right) = \frac{
C_{A_{\alpha}}^{5z}
\left(t; \boldsymbol{p}^{\prime}, \boldsymbol{p}\right)}{C_{SS}\left(t_{\mathrm{sink}}-t_{\mathrm{src}}; \boldsymbol{p}^{\prime}\right)}
\sqrt{\frac{C_{L S}\left(t_{\mathrm{sink}}-t; \boldsymbol{p}\right) C_{SS}\left(t-t_{\mathrm{src}}; \boldsymbol{p}^{\prime}\right) C_{L S}\left(t_{\mathrm{sink}}-t_{\mathrm{src}}; \boldsymbol{p}^{\prime}\right)}{C_{L S}\left(t_{\mathrm{sink}}-t; \boldsymbol{p}^{\prime}\right) C_{S S}\left(t-t_{\mathrm{src}}; \boldsymbol{p}\right) C_{LS}\left(t_{\mathrm{sink}}-t_{\mathrm{src}}; \boldsymbol{p}\right)}},
\end{align}
which yields the following asymptotic values in the asymptotic region 
($t_{\mathrm{sep}}/a\gg (t-t_{\mathrm{src}})/a \gg 1$):
\begin{align}
    \label{eq:fa_def}
    \mathcal{R}^{5z}_{A_i}(t;\boldsymbol{q})
    =
    \frac{1}{Z_A}
    \sqrt{\frac{E_N+M_N}{2E_N}}
    \left[
        {F}_A(q^2)\delta_{i3}-\frac{q_iq_3}{E_N+M_N}{F}_P(q^2)
    \right].
\end{align}
We evaluate the values of the target form factor in the standard plateau method for $F_A$, 
while our new method using time-derivative correlator is employed for $F_P$~\cite{Sasaki:2024PoSLattice},
in order to reduce leading $\pi N$ contributions.

Finally in this study,
we evaluate $g_A$ from $F_A$ at zero-momentum transfer,
while
we use the z-expansion
which in known as an model-independent $q^2$-parameterization
for $(q^2+m_\pi^2)F_P(q^2)$ to evaluate of the values of $g_P^*$ and $g_{\pi NN}$.
In order to employ the z-expansion,
the factor $(q^2+m_\pi^2)$ is involved in our analysis to factor out the pole singularities associated with the pion in the timelike region.
For a detail of the z-expansion method,
see Ref.~\cite{Hill:2010yb, Tsuji:2023llh} and all relevant references therein.

\section{Simulation details}
\label{sec:simulation_details}

We use the PACS10 configurations generated by the PACS Collaboration with the six stout-smeared ${\mathscr{O}}(a)$ improved Wilson-clover quark action and Iwasaki gauge action at $\beta=1.82$, $2.00$ and $2.20$ corresponding to the lattice spacings of $0.09$ fm (coarse), $0.06$ fm (fine) and $0.04$ fm (superfine), respectively~\cite{Shintani:2018ozy, Tsuji:2022ric, Tsuji:2023llh}. 
When we compute nucleon two-point and three-point functions, the all-mode-averaging (AMA) technique~\cite{Blum:2012uh, Shintani:2014vja, vonHippel:2016wid} is employed in order to reduce the statistical errors significantly without increasing
computational costs. 
The nucleon interpolating operators defined in Eq.~(\ref{eq:interpolating_op}) are exponentially smeared with $(A, B)=(1.2,0.16)$ for $128^4$ lattice ensemble, $(A,B)=(1.2,0.11)$ for $160^4$ lattice ensemble and $(A,B)=(1.2, 0.07)$
for $256^4$ lattice ensemble. 
As for the three-point functions, the sequential source method is employed and calculated with $t_{\rm sep}/a=\{10,12,14,16\}$ for $128^4$ lattice ensemble, $t_{\rm sep}/a=\{13,16,19\}$ for $160^4$ lattice ensemble
and $t_{\rm sep}/a=\{20,29\}$ for $256^4$ lattice ensemble.
In addition, the renormalization factor $Z_V$ and $Z_A$ are determined by the Schr\"odinger functional method (see Appendix E in Ref.~\cite{Tsuji:2023llh}).

%
%
\begin{table*}[ht]
\caption{
Parameters for three sets of the PACS10 ensembles.
See Refs.~\cite{Shintani:2018ozy, Tsuji:2022ric, Tsuji:2023llh} for further details.
\label{tab:simulation_details}}
\centering
\begin{tabular}{l|cccccccc}
\hline \hline
 & $\beta$ &$L^3\times T$ & $a^{-1}$ [GeV] &  $\kappa_{ud}$ & $\kappa_{s}$ &$c_{\mathrm{SW}}$ &  $m_\pi$ [GeV] \\
          \hline
 Superfine & 2.20& $256^3\times 256$ &4.8& 0.1254872 & 0.1249349 & 1.00 & 0.142\\
 Fine      & 2.00& $160^3\times 160$ &3.1& 0.125814  & 0.124925  & 1.02 & 0.138\\
 Coarse    & 1.82& $128^3\times 128$ &2.3& 0.126177  & 0.124902  & 1.11 & 0.135\\
\hline \hline
\end{tabular}
\end{table*}

\section{Numerical results}
\label{sec:numerical_results}
In this study,
we would like to present the preliminary results of the nucleon dispersion relation, axial-vector coupling $g_A$, the induced pseudoscalar coupling $g_P^*$ and the pion-nucleon coupling $g_{\pi NN}$ obtained by three sets of the PACS10 gauge configurations.
Although the results obtained from the superfine lattice are still very preliminary,
we will discuss the finite lattice spacing effect on these quantities,
by comparing the results obtained from all three ensembles.

\subsection{Nucleon dispersion relation}
\label{ssec:nucleon_dispersion_relation}
\begin{figure*}
\centering
\includegraphics[width=0.69\textwidth,bb=0 0 792 612,clip]{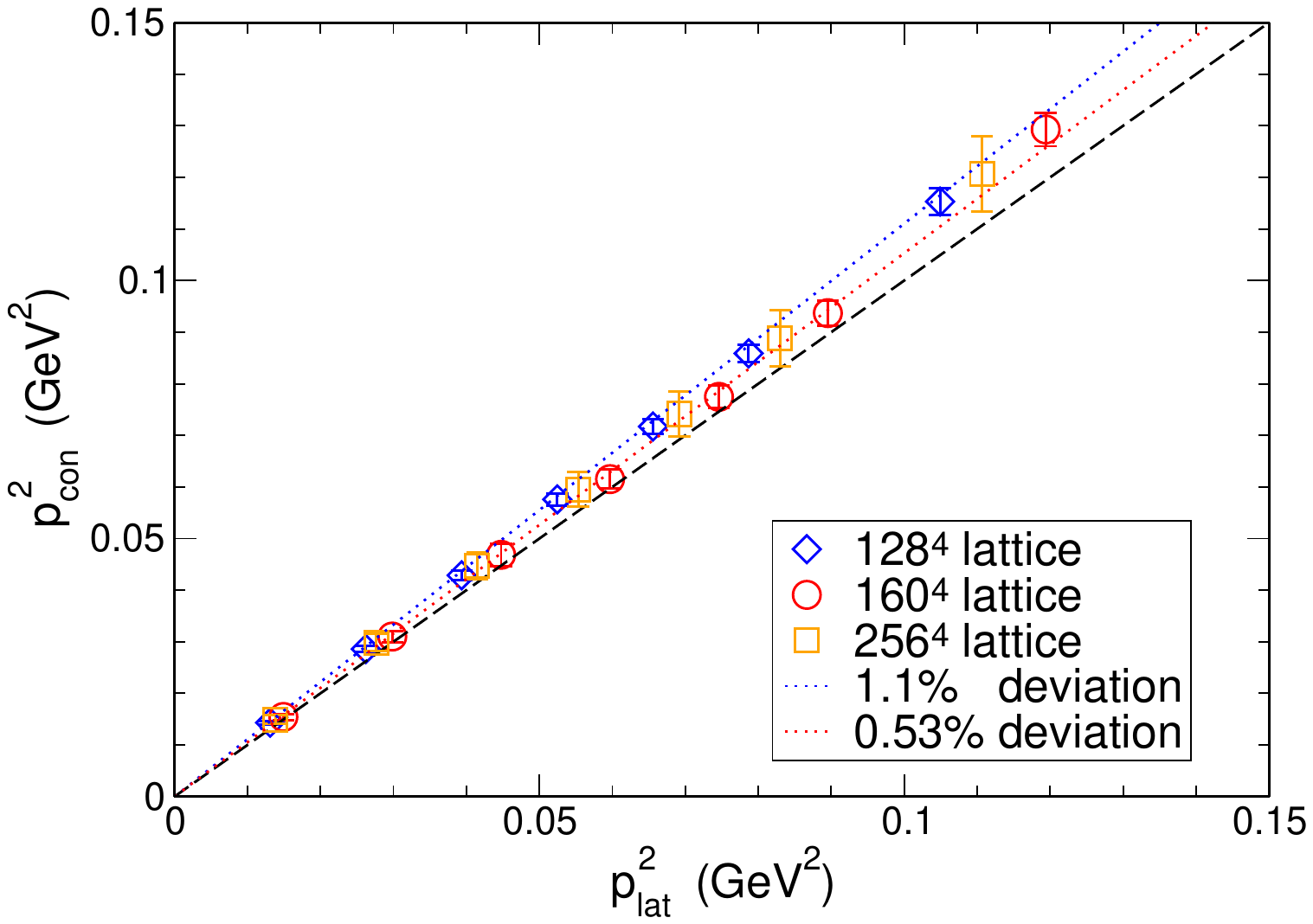}
\caption{
Check of the dispersion relation for the nucleon with an improved method described in Ref.~\cite{Tsuji:2023llh}. 
A dashed line represents the relativistic continuum dispersion relation, while red and blue dotted lines
are given by the linear fit of the coarse and fine data set, respectively.
}
\label{fig:Disp_from_DeltaE_N}
\end{figure*}

First, let us discuss the nucleon dispersion relation,
in order to
determine how much the finite lattice spacing effect is potentially present in our nonperturbatively $O(a)$-improved  Wilson-clover quark action.
Figure ~\ref{fig:Disp_from_DeltaE_N} shows the nucleon dispersion relation obtained with our three lattices.
The horizontal axis shows the momentum squared given by lattice momentum as
$p_{\mathrm{lat}}^2=\left(\frac{2\pi}{aL}\right)^2\times |\boldsymbol{n}|^2$
with $|\boldsymbol{n}|^2=1,2,\cdots,6,8$,
while the vertical axis represents the momentum squared obtained from
$p_{\mathrm{con}}^2=E_N^2(\boldsymbol{p})-M_N^2$
with the nucleon energies $E_N(\boldsymbol{p})$ and nucleon mass $M_N$ in physical units~\footnote{
For a more accurate check of the dispersion relation,
we evaluate the momentum squared as
$p_{\mathrm{con}}^2 =\Delta E_N(\boldsymbol{p})(\Delta E_N(\boldsymbol{}{p})+2 M_N)$
with the energy splittings $\Delta E_N(\boldsymbol{p}) \equiv E_N(\boldsymbol{p}) - M_N$.
For detail, see Ref.~\cite{Tsuji:2023llh}.}.
This result implies that the finite lattice spacing effect on the dispersion relation is not significant in every lattice ensemble. 
In detail,
the deviation is at most 1.1\% for the coarse lattice and 0.53\% for the fine lattice from a slope of the continuum one,
which is expected from our usage of nonperturbatively $O(a)$-improved Wilson fermions.
The preliminary results from the superfine lattice still have large statistical errors, but they are consistent with the data points
obtained from the coarse and fine lattices.

\subsection{Nucleon axial-vector, induced pseudoscalar and pion-nucleon couplings}
\label{ssec:nucleon_couplings}
\begin{figure*}
\centering
\includegraphics[width=0.32\textwidth,bb=0 0 528 612,clip]{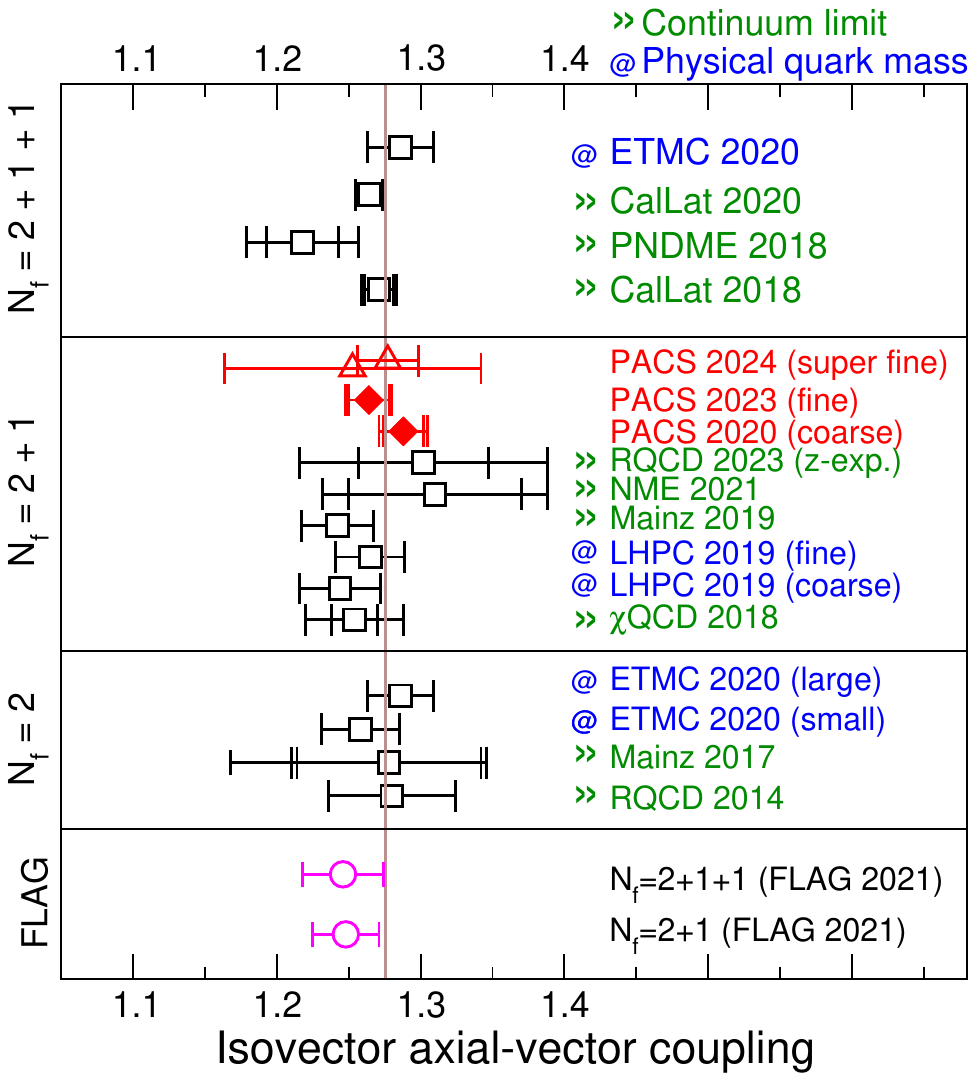}
\includegraphics[width=0.32\textwidth,bb=0 0 528 612,clip]{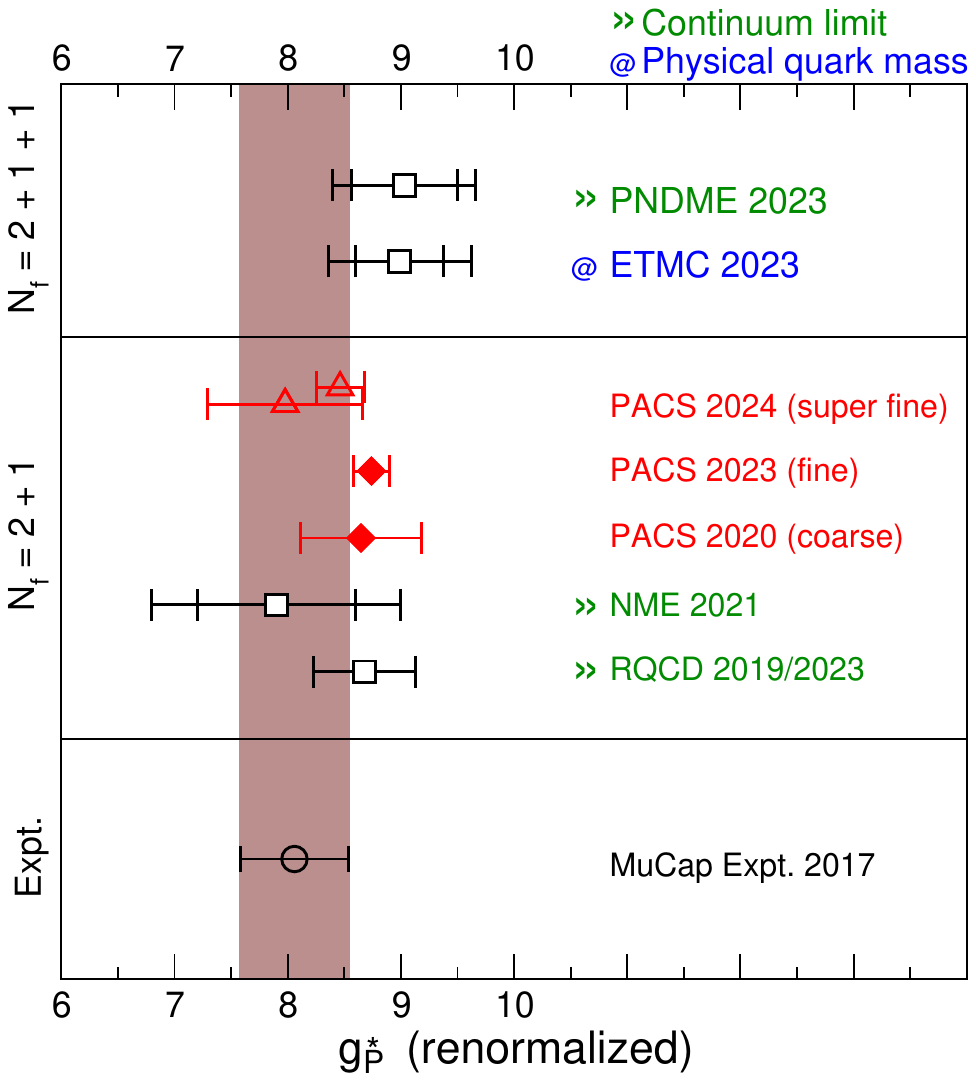}
\includegraphics[width=0.32\textwidth,bb=0 0 528 612,clip]{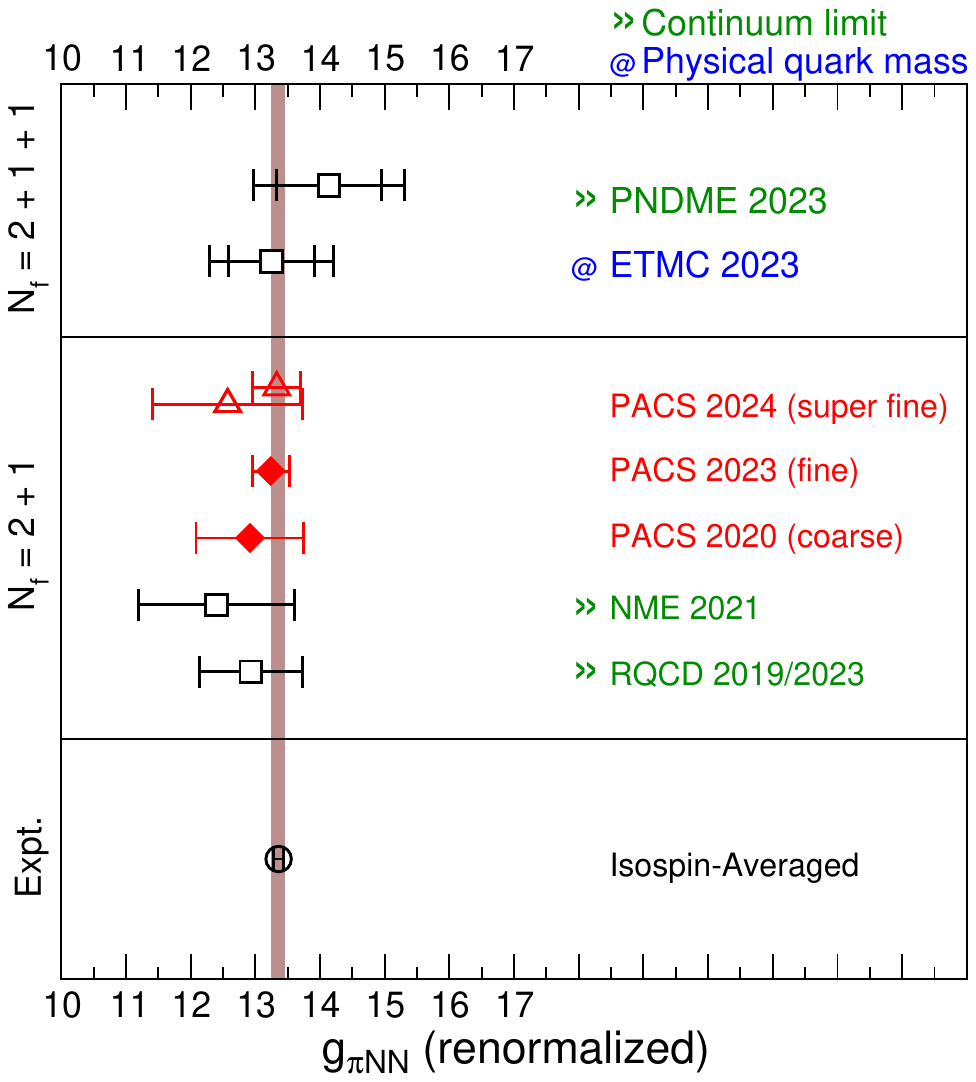}
\caption{Summary for our current status, the experimental values and other lattice QCD results
for the axial-vector coupling (left), induced pseudoscalar coupling (middle) and pion-nucleon coupling (right).
The inner error bars represent the statistical errors,
while the outer error bars are evaluated by both the statistical and systematic errors added in quadrature.
Uncertainties stemming from the excited-state contamination that is encoded as the $t_\mathrm{sep}$-dependence are taken into account in the systematic errors.
Two open triangle symbols represent the preliminary results obtained from the superfine lattice with $t_\mathrm{sep}=0.8$ fm (the upper value with the smaller error bar) and $t_\mathrm{sep}=1.2$ fm (the lower value with the larger error) fm, respectively.
The experimental values denoted as a brown band in each panel.
The values of $g_A$ and $g_P^*$ are taken from Ref.~\cite{ParticleDataGroup:2022pth} and \cite{MuCap:2015boo},
while the value for $g_{\pi NN}$ shown here is given by the isospin average
as described in the text.
}
\label{fig:comparison}
\end{figure*}

Figure \ref{fig:comparison} shows summary plots
for our current status of three couplings, together with
the experimental values and the other lattice QCD results.
The inner and outer error bars represent the statistical uncertainties
and the total uncertainties that are given by adding the statistical
errors and systematic errors in quadrature,
while
only the statistical error is shown for our preliminary results obtained from the superfine lattice denoted as the open triangle symbols.
The systematic errors take into account
uncertainties stemming from the excited-state contamination and so on
(see details in \textbf{}~Ref.~\cite{Tsuji:2023llh}).

Let us first discuss the results for the axial-vector coupling $g_A$
which is precisely measured by the experiments. 
In the left panel of Fig.~\ref{fig:comparison}, 
we show the results of $g_A$.
Our results of $g_A$ are evaluated with the axial form factor $F_A$ at the zero-momentum transfer renormalized with $Z_A$, which is described in Eq.~(\ref{eq:couplings}).
Our results obtained at three different lattice spacings including the superfine lattice can reproduce the experimental value within statistical precision.
The systematic uncertainty of the excited-state contamination is examined for
the results with the coarse and fine lattices
by the combined analysis with the range of $1.0 \lesssim t_\mathrm{sep}[\mathrm{fm}] \lesssim 1.2$, while
for the preliminary results of the superfine lattice,
individual data obtained with $t_\mathrm{sep}=\{0.8,1.2\}$ fm
are plotted without the combined analysis.
Hence we do not have any firm conclusion based on the continuum-limit extrapolation, while
no significant dependence on the lattice spacing is observed 
within the current statistical uncertainties.
The continuum-limit extrapolation requires a further calculation for the superfine lattice,
and all of the systematic uncertainties including the excited-state contamination should be examined in future.

The middle and right panels of Fig.~\ref{fig:comparison} represent the results for the induced pseudoscalar coupling $g_P^*$ and the pion-nucleon coupling $g_{\pi NN}$, respectively.
In order to evaluate these coupling,
we calculate the values of $F_P(q^2)$ at 7 points of the momentum transfer in the range of
$0.014 \lesssim q^2[\mathrm{GeV}^2] \lesssim 0.11$,
and parametrize the $q^2$-dependence with the z-expansion method.
As for the pion-nucleon coupling,
the experimental value shown in the right panel of Fig.~\ref{fig:comparison} are evaluated as an isospin-averaged value~\footnote{The isospin average for the pion-pion nucleon coupling is given by
$
g_{\pi NN}^2
=
\frac{1}{3}
(g_{\pi^0 NN}^2 + 2 g_{\pi^\pm NN}^2)
$
.}
evaluated with the experimental data for charged and neutral pions~\cite{Babenko:2016idp, Limkaisang:2001yz},
in order to compare with the lattice QCD results,
where both QED and isospin breaking effects are not taken into account.
The errors on our results of $g_P^{\ast}$ and $g_{\pi NN}$
are evaluated in the same way as $g_A$. 
Our obtained values are in good agreement with the other lattice QCD results and the pion-pole dominance model,
both of which reproduce the experimental values.

We emphasize here that the excited-state contamination has been a long-standing obstacle to the accurate calculation of the two couplings associated with the induced pseudoscalar form factor $F_P$. However, in our calculations where careful tuning of the smearing parameters for
the nucleon ground state was performed, we have succeeded in completely removing the leading $\pi N$ contribution from the analysis of the $F_P$
data by our newly proposed method~\cite{Sasaki:2024PoSLattice}.
As for the lattice discretization uncertainties,
no significant dependence on the lattice spacing is observed as well as the nucleon axial-vector coupling.
However, it is still not precise enough to take the continuum limit in order to draw firm conclusions.

%
%
%

%
%

\section{Summary}
\label{sec:summary}

We calculate 
the nucleon dispersion relation, axial-vector coupling $g_A$, the induced pseudoscalar coupling $g_P^*$ and the pion-nucleon coupling $g_{\pi NN}$
using three sets of the PACS10 ensembles generated at the physical point on a $(10\;{\mathrm{fm}})^4$ volume.
The PACS10 gauge configurations are generated by the PACS Collaboration with the stout-smeared $O(a)$ improved Wilson quark action and Iwasaki 
gauge action.

Since the results obtained from the superfine lattice is still very preliminary,
the continuum-limit extrapolation has not been performed yet.
However, according to the nucleon dispersion relation, we observed that the finite lattice spacing effect is not so large, with the deviation being at most $1.1\%$ even for the coarse lattice, as expected from our improved action.
In addition,
it was found that there is no strong dependence on the lattice spacing in our results for the three couplings,
and all three quantities are consistent with the corresponding results obtained from the other lattice QCD simulations and in good agreement with the corresponding experimental values within their errors.
In addition, it is important to emphasize that
our new method that can remove the leading $\pi N$ contribution from the $F_P$ form factor works well and helps to accurately determine the induced pseudoscalar coupling $g_P^{\ast}$ and pion-nucleon coupling $g_{\pi NN}$.

Needless to say that a comprehensive study of the discretization uncertainties and also the continuum limit extrapolation of the target quantities require additional lattice simulations with the superfine lattice to reduce both statistical and systematic uncertainties.
Further calculations using the superfine PACS10 ensemble is now underway.

\section*{Acknowledgement}
We would like to thank members of the PACS collaboration for useful discussions.
K.-I.~I. is supported in part by MEXT as ``Feasibility studies for the next-generation computing infrastructure".
Numerical calculations in this work were performed on Oakforest-PACS in Joint Center for Advanced High Performance Computing (JCAHPC) and Cygnus  and Pegasus in Center for Computational Sciences at University of Tsukuba under Multidisciplinary Cooperative Research Program of Center for Computational Sciences, University of Tsukuba, and Wisteria/BDEC-01 in the Information Technology Center, the University of Tokyo. 
This research also used computational resources of the K computer (Project ID: hp1810126) and the Supercomputer Fugaku (Project ID: hp20018, hp210088, hp230007, hp2320199, hp240207) provided by RIKEN Center for Computational Science (R-CCS), as well as Oakforest-PACS (Project ID: hp170022, hp180051, hp180072, hp190025, hp190081, hp200062),  Wisteria/BDEC-01 Odyssey (Project ID: hp220050) provided by the Information Technology Center of the University of Tokyo / JCAHPC.
The  calculation employed OpenQCD system(http://luscher.web.cern.ch/luscher/openQCD/). 
This work is supported by the JLDG constructed over the SINET5 of NII.
This work was also supported in part by Grants-in-Aid for Scientific Research from the Ministry of Education, Culture, Sports, Science and Technology (Nos. 18K03605, 19H01892, 22K03612, 23H01195, 23K03428, 23K25891) and MEXT as ``Program for Promoting Researches on the Supercomputer Fugaku'' (Search for physics beyond the standard model using large-scale lattice QCD simulation and development of AI technology toward next-generation lattice QCD; Grant Number JPMXP1020230409).

\bibliographystyle{man-apsrev2}
\bibliography{skeleton}

\end{document}